# Synthesis and characterization of vertically oriented hybrid $Zn_2GeO_4$-ZnO beaded nanowire arrays and $Zn_2GeO_4$ nanotubes

Bablu Mukherjee, Binni Varghese, Minrui Zheng, Suzi Deng, Eng Soon Tok, Chorng Haur Sow

Department of Physics, 2 Science Drive 3, National University of Singapore (NUS), Singapore-117542

Vertically aligned, beaded zinc germinate ($Zn_2GeO_4$)/ zinc oxide (ZnO) hybrid nanowire arrays were successfully synthesized by a vapor-solid process via a catalyst-free approach. The as-synthesized products were characterized using X-ray diffraction, scanning electron microscopy and transmission electron microscopy equipped with an energy-dispersive X-ray spectrometer. TEM studies revealed the beaded microstructures of the $Zn_2GeO_4$/ZnO nanowire. Furthermore, $Zn_2GeO_4$ nanotubes were synthesized after wet etching treatments on $Zn_2GeO_4$/ZnO hybrid nanowires.

## 1. Introduction

One dimensional (1D) nanostructure, such as nanowires, nanotubes and nanobelts, of ternary metal oxides often shows interesting properties and functionalities beyond their simple binary metal oxide systems [1]. Among the many ternary systems $Zn_2GeO_4$ has interesting properties which make it suitable for applications such as visible-blind deep-ultraviolet photodetection [2, 3], high-capacity anode material of lithium battery [4], bright white-bluish luminescence [5], photocatalytic water-splitting [6, 7], photocatalytic reduction of $CO_2$ into renewable hydrocarbon fuel [8]. However, the synthesis of 1D ternary nanostructures with pure phase and desirable stoichiometry is considerably more challenging in comparison with the synthesis of 1D structures of their binary counterparts. Here we have demonstrated the control synthesis of 1D hybrid $Zn_2GeO_4$/ZnO nanowires array using aligned ZnO nanowires sample as a reactive template. The growth of $Zn_2GeO_4$ beads on ZnO nanowire and the in-situ annealing was performed together. During high-temperature growth and annealing, the interplay of Ge into ZnO matrix most likely facilitates the growth of $Zn_2GeO_4$ ternary crystallites and form beaded hybrid $Zn_2GeO_4$/ZnO nanowire.

## 2. Experimental

The synthesis of hybrid $Zn_2GeO_4$/ZnO nanowires array was performed in a two-step process. Firstly the primary vertically aligned ZnO nanowire sample was synthesized on Si substrate with pre-deposited ZnO thin seed layer and then the as grown ZnO nanowire sample was used as a reactive template to grow hybrid $Zn_2GeO_4$/ZnO nanowire. Aligned ZnO nanowire was synthesized in a horizontal tube furnace via a catalyst free approach as reported earlier [9]. Then the ZnO nanowire substrate was placed near to a small piece of Ge wafer for the hybrid nanowire growth. The Ge wafer was used as the source material. A small one end closed quartz tube containing ZnO nanowire substrate and Ge wafer was inserted into the big ceramic tube such that the ZnO sample and Ge wafer will be exposed in the middle of uniform heating zone of the tube furnace. Purified Ar gas was flowed with a total flow rate of 100 sccm (standard cubic centimeter per minute) through the vacuum sealed tube furnace for 60 mins before ramping the temperature to 850 $^{\circ}$C at a rate of 15 $^{\circ}$C/min. The synthesis of $Zn_2GeO_4$/ZnO nanowires was conducted at 850 $^{\circ}$C for 60 mins. Inside pressure of the furnace was maintained at fixed value of 1 mbar during the whole synthesis process.

## 3. Results and Discussion

**Fig. 1 (a-b)** shows the scanning electron microscopy (SEM) images of aligned ZnO nanowires and hybrid $Zn_2GeO_4$/ZnO nanowires before and after high-temperature annealing treatment, respectively. Vertically aligned ZnO nanowires (**Fig. 1a**) with hexagonal side faced have diameters ranging from 60 nm to 100 nm and lengths around 1-2 μm. The microstructured beads along the hybrid nanowire can be clearly viewed from the tilted view of the SEM image, **Fig. 1(c)**. The SEM images with 20 degree tilted view and different magnifications of the beaded $Zn_2GeO_4$/ZnO hybrid nanowires are shown in **Fig. 1(c-d)**, which shows that the hybrid nanowires are still oriented.

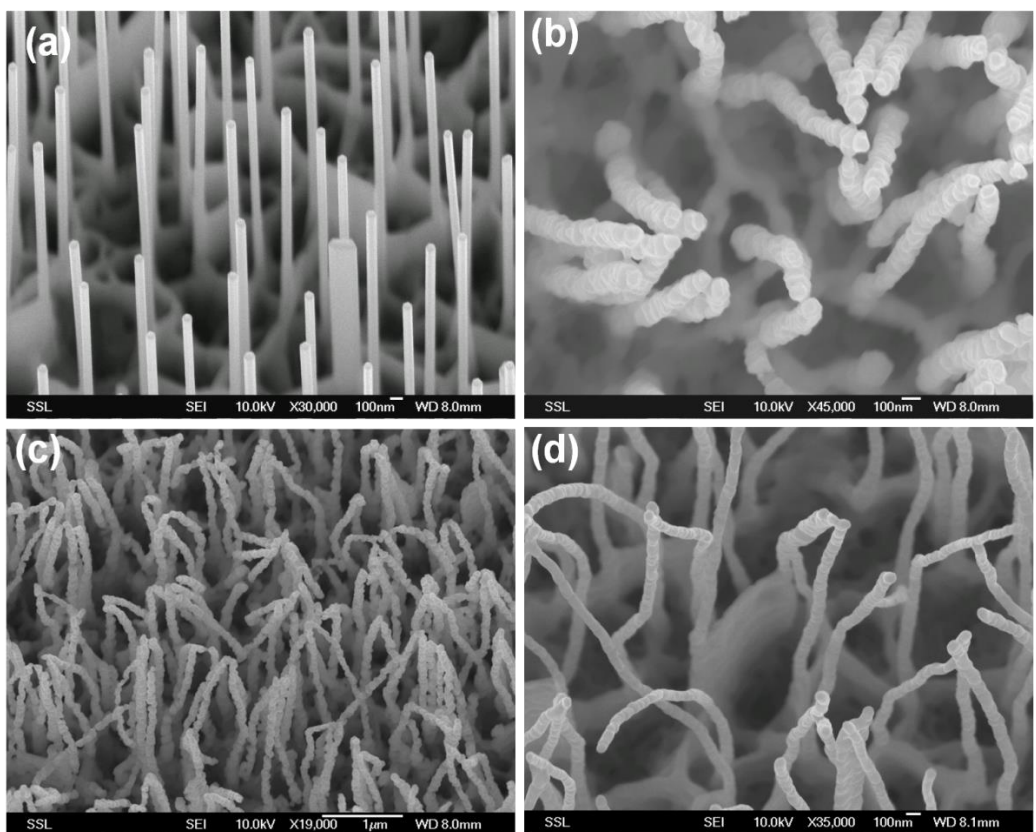

**Fig. 1.** SEM images of (a) pristine ZnO nanowires (tilted view: 20$^{\circ}$), (b) top view of hybrid $Zn_2GeO_4$/ZnO aligned nanowires after growth at 850 $^{\circ}$C for 60 mins, (c) tilted view (20$^{\circ}$) of hybrid nanowires and (d) higher magnification image of hybrid nanowires.

To obtain more details about the crystal structures and compositions of the final product, TEM, selected-area electron diffraction (SAED) and energy dispersive X-ray spectroscopy (EDS) measurements equipped with TEM were performed. The TEM image shown in **Fig. 2(a,b)** reveals that the as-synthesized hybrid $Zn_2GeO_4$/ZnO nanowires are beaded with diameter ranging from 60 nm to 100 nm. **Fig. 2(b)** indicates the presence of planer defects in the region between two adjacent beads. The inset in **Fig. 2(b)** shows the selected area electron diffraction (SAED) pattern of the $Zn_2GeO_4$ structure. The lattice spacing between the adjacent lattice planes (**Fig. 2(c)**) is measured to be approximately 0.712 nm, which is consistent with the (110) planer separation of $Zn_2GeO_4$ crystal, confirming the direction of growth of the nanobeads to be along [110] direction. The fast Fourier transform (FFT) pattern inserted in the bottom-right corner of **Fig. 2(c),** which matched with the FFT of $Zn_2GeO_4$ (110) plane. EDS spectrum (**Fig. 2(d)**) shows that the beaded nanowires consist of only three elements, namely, Zinc, Germanium and Oxygen with atomic percentage of 29.8, 14.9 and 55.3, respectively. This is approximately in the ratio of 2: 1: 4, consistent with the stoichiometry of $Zn_2GeO_4$ compound.

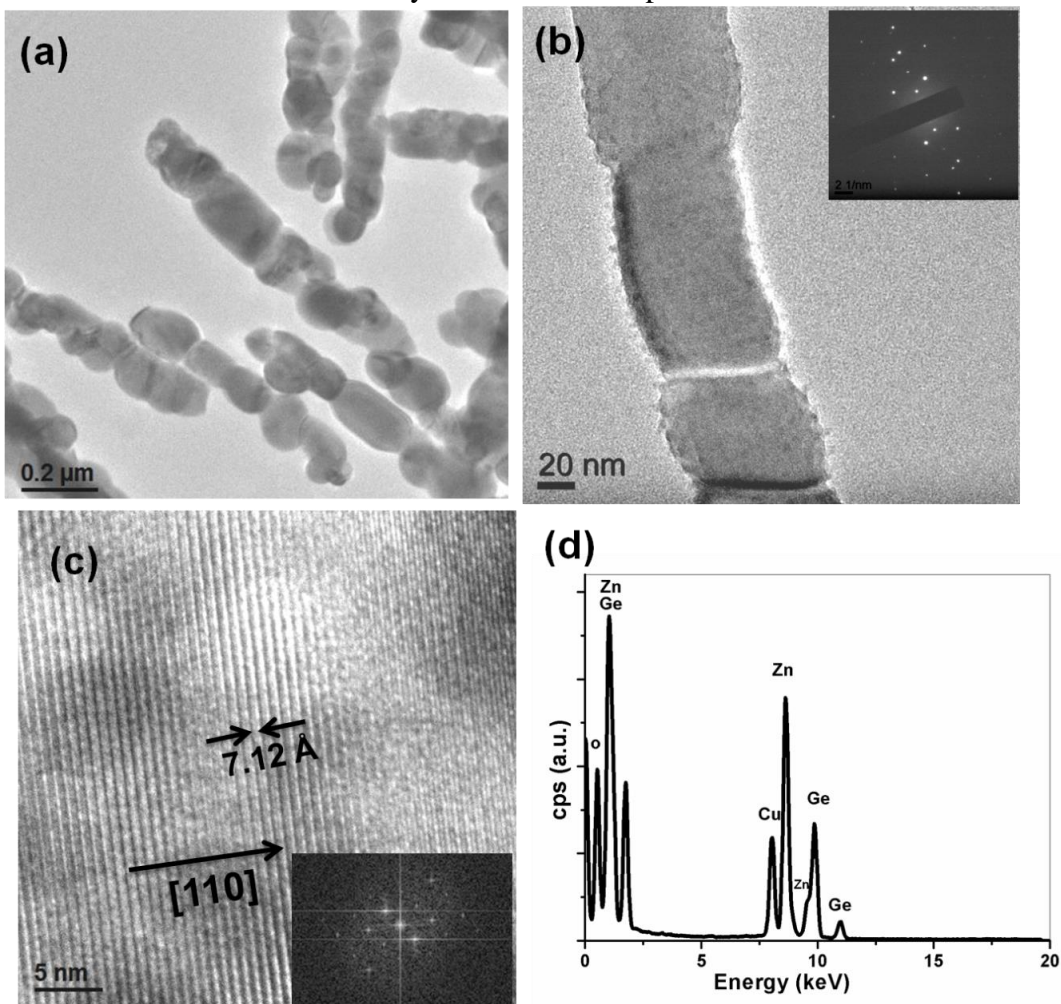

**Fig. 2.** Bright–field TEM images of **(a)** $Zn_2GeO_4$/ZnO beaded nanowires, **(b)** a single nanowire and inset shows a SAED pattern. **(c)** High-resolution TEM image, confirming the <110> growth direction of the $Zn_2GeO_4$ epitaxial with c-oriented ZnO nanowire; the inset picture corresponds to the FFT transformation. **(d)** EDS spectra of the $Zn_2GeO_4$/ZnO beaded nanowires.

The wet etching treatments on $Zn_2GeO_4$-ZnO beaded nanowires were done in the following way. The nanowires substrate was rinsed with deionized water and transferred into 0.04 M ethylenediamine (EDA) aqueous solutions and heated at 70 °C for a period of 12 hours to etch the core ZnO. The obtained product was rinsed with deionized water and ethanol several times and then dried with $N_2$ between rinses. The morphology of the obtained product is shown in **Fig. 3.** The hollow interior of beaded nanotubes can be seen from the SEM images **Fig. 3 (a,b)**.

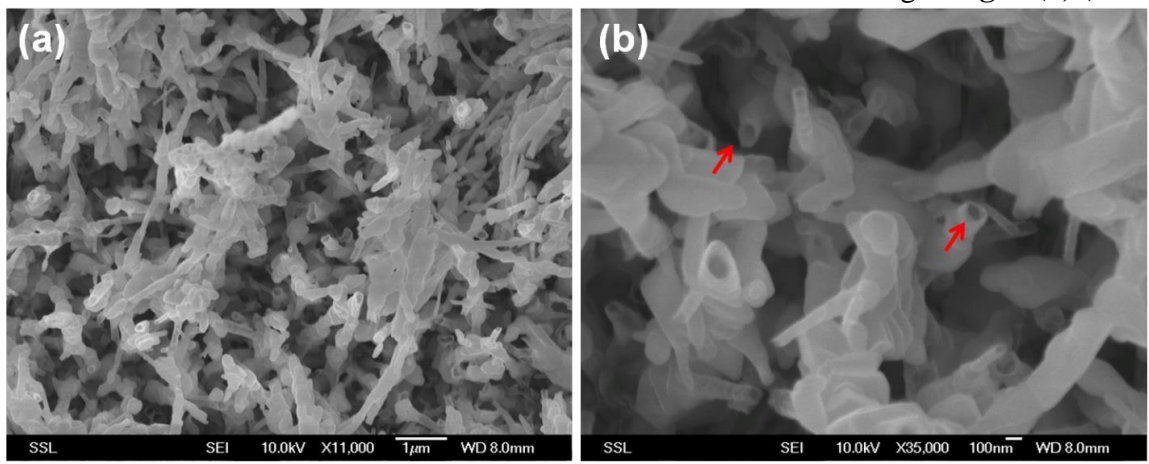

**Fig. 3.** SEM images of **(a)** tubular beaded $Zn_2GeO_4$ structures after a partially removal of the unconsumed ZnO core, and **(b)** the hollow interior of $Zn_2GeO_4$ nanotubes are observed as marked with the red arrows.

X-ray diffraction (XRD) analysis was performed to know the crystal phase of ZnO nanowires and hybrid $Zn_2GeO_4$/ZnO nanowires. **Fig. 4.** shows the XRD patterns of pristine ZnO nanowires (curve a), hybrid $Zn_2GeO_4$/ZnO nanowires (curve b) and $Zn_2GeO_4$ nanotubes after wet etching of unconsumed ZnO core (curve c) for comparison. In 'curve a' all peaks can be easily indexed with wurtzite ZnO crystal with lattice constants of a = 3.25 Å, c = 5.21 Å [JCPDS Card No. 80-0075]. For hybrid $Zn_2GeO_4$/ZnO nanowires, all the peaks in 'curve b' can be indexed as a mixture of rhombohedral zinc germanium oxide ($Zn_2GeO_4$) crystal with lattice constants of a = 14.23 Å, c = 9.53 Å [JCPDS: PDF File No. 00-011-0687] and wurtzite ZnO crystal (JCPDS: PDF File No. 01-070-2551) with one additional peak of Si (110) substrate. It can be seen from 'Curve c' that the peaks intensities corresponding to ZnO have been much reduced after wet chemical etching of hybrid $Zn_2GeO_4$/ZnO nanowires.

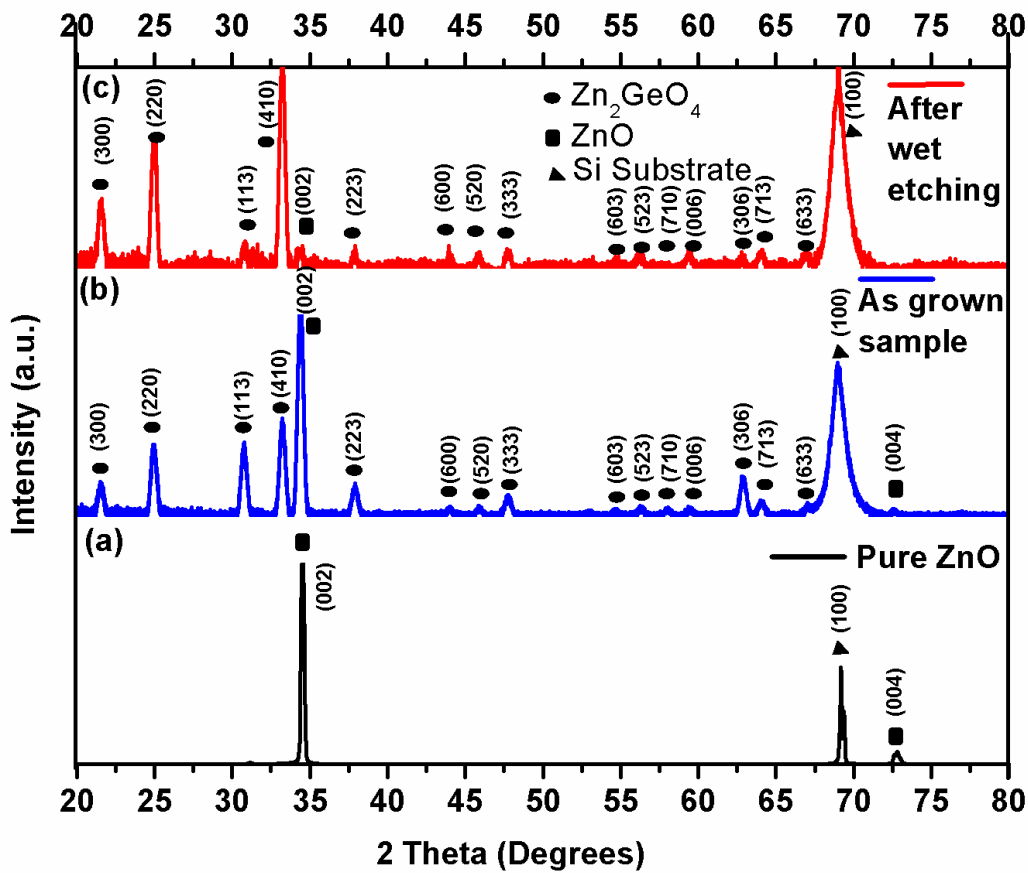

**Fig. 4.** XRD pattern recorded from (a) pure vertically aligned ZnO nanowire sample, (b) the as grown $Zn_2GeO_4$/ZnO beaded nanowire sample, and (c) the $Zn_2GeO_4$ beaded hollow nanotube sample after wet chemical etching of hybrid $Zn_2GeO_4$/ZnO nanowire for 20 hours at 80 $^o$C.

## 4. Conclusions

In conclusion, aligned hybrid $Zn_2GeO_4$-ZnO heterostructures have been synthesized via a high temperature vapor-solid process using pristine vertically oriented ZnO nanowires as the template. The surface morphology, microstructures, chemical composition and crystal structures were examined through SEM, TEM, EDS and XRD. We have demonstrated the uses of ZnO nanowires for hybrid $Zn_2GeO_4$-ZnO nanowires growth. Conventionally, we discussed the synthesis of $Zn_2GeO_4$ nanotubes after wet chemical etching of hybrid $Zn_2GeO_4$/ZnO nanowires. It is expected that such hybrid $Zn_2GeO_4$-ZnO nanowire should support as the backbone for ZnO nanobranches growth, which will effectively produce three dimensional (3D) heterostructures of $Zn_2GeO_4$-ZnO nanowires. These hybrid materials might have application in supercapacitor and in high-capacity anode material for lithium batteries.